\documentclass{aa}

\usepackage[dvips]{graphicx}
\usepackage{txfonts}
\usepackage{natbib}
\usepackage[latin1]{inputenc}
\newcommand{\bvec}[1]{\mbox{\boldmath$#1$} }  
\newcommand{\bnabla}{\mbox{\boldmath$\nabla$} } 

\usepackage[usenames]{color}

\begin{document}

\title{Structure and evolution of solar supergranulation using SDO/HMI data}

\author{ Th.~Roudier\inst{1}, M.~\v{S}vanda\inst{2,3} M.~Rieutord,\inst{1},
J.M.~Malherbe\inst{4}, R.~Burston\inst{5}, L.~Gizon\inst{5,6}}

\date{Received \today  / Submitted }

\offprints{Th. Roudier}
 
\institute{Institut de Recherche en Astrophysique et Plan\'etologie, Universit\'e de Toulouse, CNRS,
14 avenue Edouard Belin, 31400 Toulouse, France
\and Astronomical Institute, Faculty of Mathematics and Physics, Charles University in Prague, V Hole\v{s}ovi\v{c}k\'{a}ch 2, 
CZ-18000, Prague~8, Czech Republic 
\and Astronomical Institute, Academy of Sciences of the Czech Republic (v. v. i.), Fri\v{c}ova~298, CZ-25165, Ond\v{r}ejov, Czech Republic
\and LESIA, Observatoire de Paris, Section de Meudon, 92195 Meudon, France
\and Max-Planck-Institut f\"ur Sonnensystemforschung, Justus-von-Liebig-Weg 3, 
37077 G\"ottingen, Germany
\and Institut f\"ur Astrophysik, Georg-August-Universit\"at G\"ottingen, 
Friedrich-Hund-Platz 1, 37077 G\"ottingen, Germany}

\authorrunning{Roudier et al.}
\titlerunning{Supergranulation study from SDO/HMI data}

\abstract{
 Studying the motions on the solar surface is fundamental for understanding how 
turbulent convection transports energy and how magnetic fields are distributed across
the solar surface.
}{
From horizontal velocity measurements all over the visible  disc of the 
Sun and using data from the Solar Dynamics Observatory/Helioseismic and Magnetic Imager (SDO/HMI), 
we investigate the structure and evolution of solar supergranulation.
}{
Horizontal velocity fields were measured by following the proper motions of
solar granules using a newly developed version of the coherent structure
tracking (CST) code. With this tool, maps of horizontal divergence were
computed. We then  segmented and identified supergranular cells and followed
their histories by using spatio-temporal labelling. With this dataset
we derived the fundamental properties of supergranulation, including
their motion.
}{
We find values of the fundamental parameters of supergranulation similar
to previous studies: a mean lifetime of 1.5 days and a mean diameter of
25~Mm. The tracking of individual supergranular cells reveals the solar
differential rotation and a poleward circulation trend of the meridional
flow. The shape of the derived differential rotation and meridional flow
does not depend on the cell size. If there is a background magnetic
field, the diverging flows in supergranules are weaker.
}{
This study confirms that supergranules are suitable tracers that
may be used to investigate the large-scale flows of the solar convection
as long as they are detectable enough on the surface.}

\keywords{The Sun: Atmosphere -- The Sun: Supergranulation -- The Sun: Convection}

\maketitle

\section{Introduction}

The dynamics of the solar surface is still a main subject of
research in solar physics. It is essential to understand the mechanisms
at the origin of the solar magnetic fields and their redistribution  
to constrain the dynamo process or the solar wind generation.
Indeed, the surface field patterns ultimately determine the coronal
field topology and the photospheric footpoints of open field lines
of the heliosphere \citep{DET04}. A recently published scenario of the
origin of the solar wind in coronal holes \citep{YANG2013} suggests that
motions driven by supergranulation advect meso-scale closed loops toward
supergranular borders. The horizontally moving loops impinge on an open
funnel that is located at the supergranular cell boundary.  This collision may
trigger magnetic reconnection,which finally forms the nascent solar wind.

\begin{figure}[!t]
\resizebox{4.0cm}{!}{\includegraphics{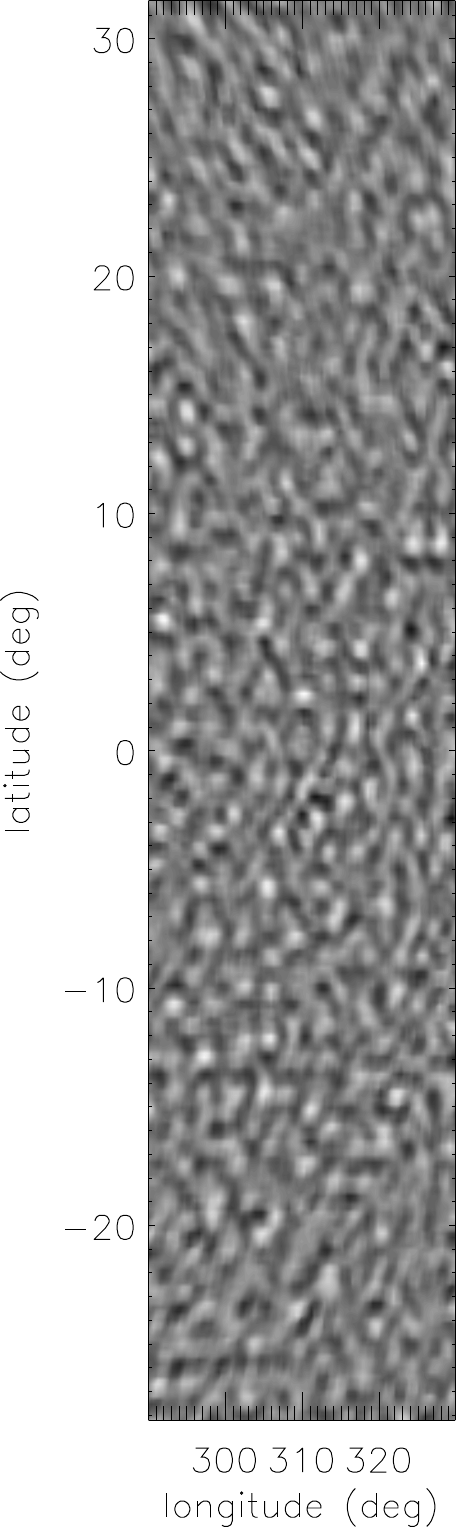}} \resizebox{4.0cm}{!}{\includegraphics{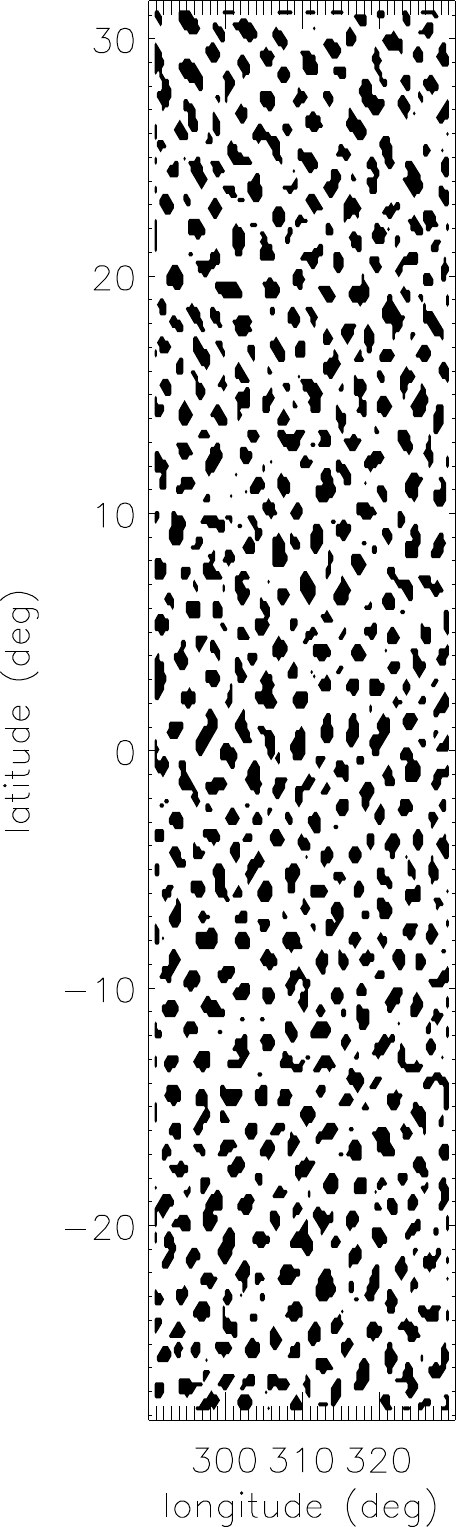}}
\caption[]{Horizontal divergence map White stands for positive divergence (left) and 
segmented divergence centres, black for positive divergence (right).}
\label{original1}
\end{figure}

Thanks to spatial observatories (SOHO, SDO, Hinode, etc.), the Sun
is the only star where large-scale (differential rotation, meridional
circulation,etc.) and small-scales (supergranulation, granulation)
flows can be observed continuously,  with temporal scales of 
a few seconds to several months. Solar granulation is generally considered
to be well understood \citep{NSA2009}, but the origin of supergranulation
is currently unknown despite of many  observations and analyses \citep{RR10}.

\begin{figure*}
\centering
\includegraphics[width=18cm]{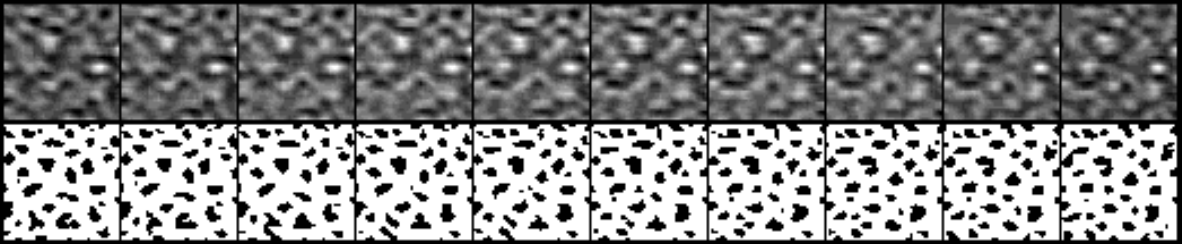}
\caption[]{Evolution of divergence: positive divergence in white (top), and the segmented 
counterpart, positive divergence in black (bottom). The time
step is 30~min and the field of view of $179\times179$~arcsec$^2$. }
\label{segtime}
\end{figure*}

Different approaches  have previously been used to study the
supergranulation: correlation or tracking techniques applied to
quiet-Sun images of CaII-K that record the emission of the magnetic
elements in the magnetic network taht surrounds the supergranular cells
\citep{DMGB07}, Doppler features \citep[][and others]{DET04,SKSB08},
or pattern of horizontal divergences of the smaller-scale flows
\citep{GDS03}. Additional information about supergranulation was also
obtained from the power spectral analyses \citep[e.g.][]{WPBL13}.

Helioseismology represents a powerful tool for studying supergranules and the
connected flows \citep[e.g.][]{Svanda2013}. The propagation of acoustic
waves through the solar interior is affected by plasma motions,
which in turn affects helioseismic observables such as frequencies or
travel times of the waves. Inverse modelling can image the flows, 
but substantial time averaging is required to obtain results with a
signal-to-noise ratio higher than unity.

The main goal of our study is to determine the properties of the
supergranulation by using horizontal flow fields directly measured on
the solar surface via solar granulation tracking computed by the coherent
structure tracking (CST) code \citep[][]{RRPM13,RRRD07, TRMR07}.

 The CST method allows determining plasma flows from scales
as small as 2.5~Mm \citep{RRLNS01} up to almost the full disc (to be precise, 
up to 0.8~radius) of the Sun. Here we take advantage of the high
spatial and temporal resolution of SDO/HMI continuum observations and
use the granulation as tracers for direct measurements of the plasma flow.

The velocity field obtained by the CST code is represented on a wavelet
basis that allows  computating the field derivatives such as divergence
or vertical vorticity.

The paper is organised in the following way: in Section 2
we explain the data reduction to isolate individual supergranules and
their histories in the datacube. In Section 3 we describe the
results obtained after applying the method. Conclusions and future 
perspectives follow.

\begin{figure*}
\sidecaption
\resizebox{7.0cm}{!}{\includegraphics{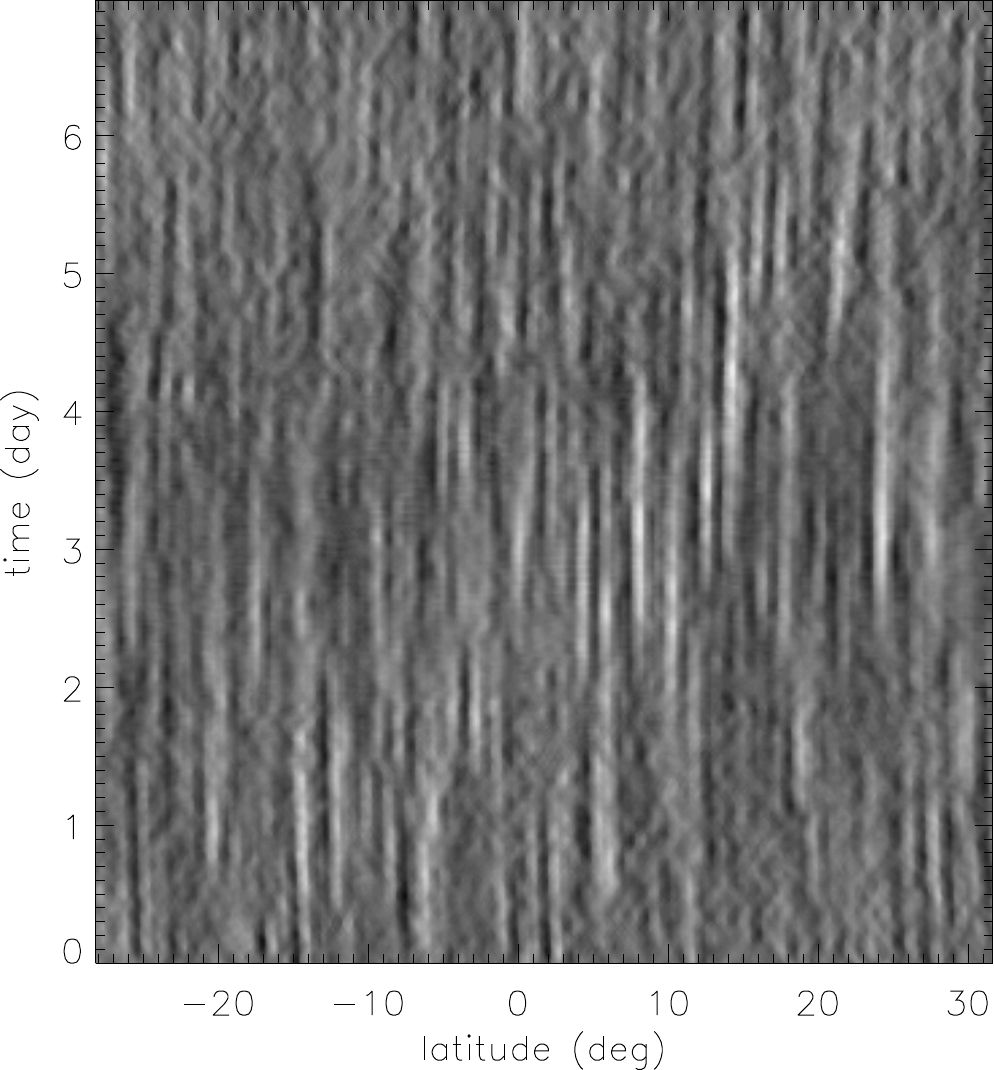}}
\resizebox{7.0cm}{!}{\includegraphics{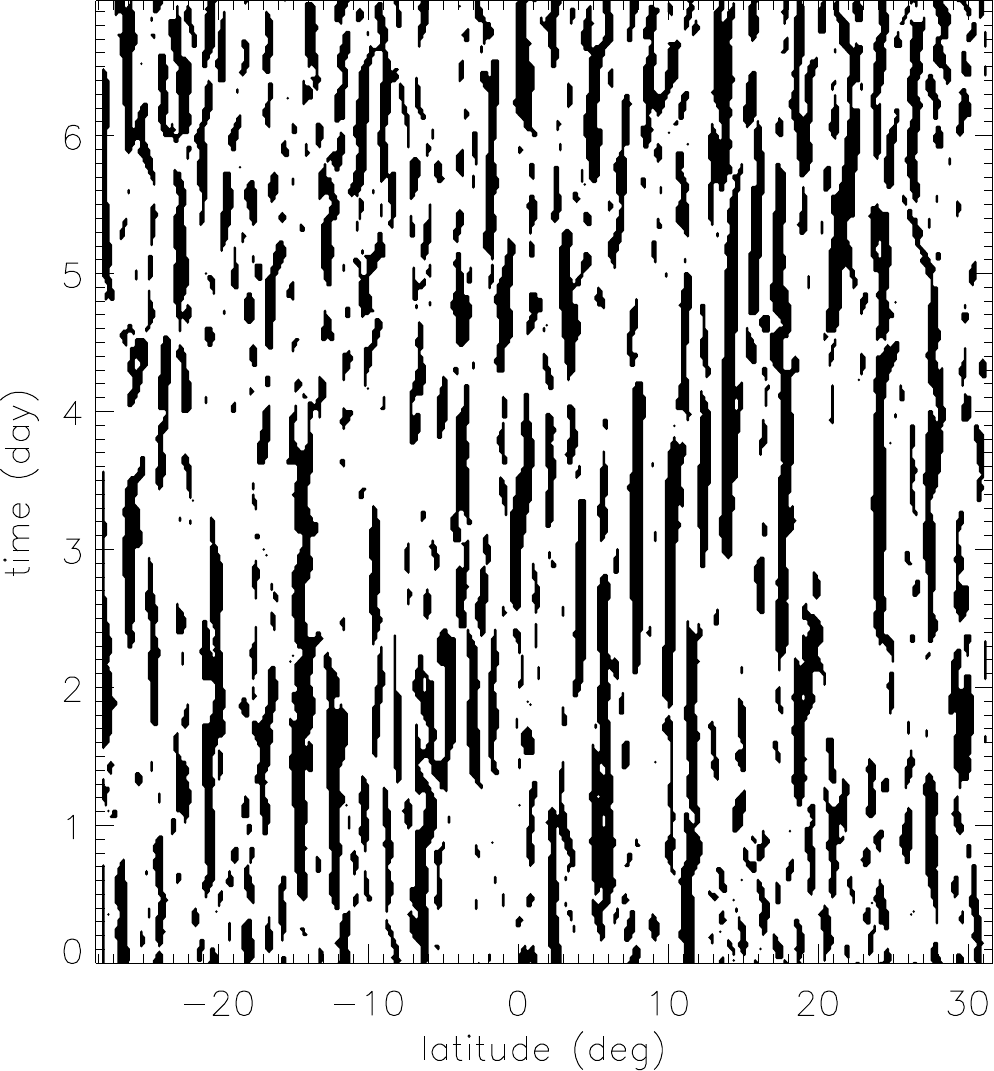}}
\caption[]{Time--latitude slice of the horizontal divergence tracked at constant rotation
rate (left) and the segmented slice (right). Linear vertical structures represent
persistent outflow cells, with their lengths proportional to their lifetimes.}
\label{original2}
\end{figure*}

\section{SDO-HMI observations and horizontal flow field measurement}

The Helioseismic and Magnetic Imager onboard the Solar Dynamics
Observatory \citep[HMI/SDO;][]{Schou2012} supplies uninterrupted
high-resolution observations of the full solar disc. This
provides a unique opportunity for mapping surface flows on
a wide range of scales (both spatial and temporal).

We used HMI/SDO white-light data from May 10 2011 to May 16 2011 (seven days 
without interruption) to derive horizontal velocity fields using
a recent version of the CST algorithm \citep{RRPM13}. The code tracked
the motion of individual granules that were recorded in the frames with a time
step of 45 seconds and a pixel size of 0.5 arcsec.

The resulting horizontal velocity fields returned by CST map the
motions of the solar surface plasma on scales larger than 2.5~Mm over
almost the entire visible surface of the Sun with a time step
of 30~minutes \citep{RRPM13}. The reconstrution of the continuous
velocity field is based on a wavelet multi-resolution analysis, by which
one can obtain velocity fields at different scales. It also gives its
derivatives such as the horizontal divergence or the vertical component
of the vorticity. The algorithm also limits the propagation of noise and
other errors \citep{RRRD07}. The measurements of horizontal flows
by CST is strictly sensitive to the motions within the thin shell at the
very top of the solar convection zone \citep{RRMV99}. It is thus suitable
for studies of solar supergranulation, especially in the region around
the disc centre, whereas local helioseismology methods are sensitive to
deeper layers and sample a thicker shell (at least 1~Mm) at the top of
the convection zone.  It was shown that the results from ~both
~methods~ are~ comparable,~ but not identical \citep{SRRBG13}.

It was described in earlier works \citep[e.g.][]{BD01,GD03,HGSD08}
that a convenient way of identifying the individual supergranular cells 
is through the use of the horizontal divergence $\bnabla_{\rm
h}\cdot\bvec{v}$ of the flow,

\begin{equation}
\bnabla_{\rm h}\cdot\bvec{v}=\partial v_x/\partial x+\partial v_y/\partial y,
\end{equation}
where $v_x$ and $v_y$ are the horizontal components of the flow field in
a local Cartesian coordinate system. The regions of positive divergence
aretypically regions of upflows, while regions with a negative horizontal
divergence depict downflows. A supergranule in a flow map near the solar
surface is defined as a compact region of positive divergence surrounded
by lanes of negative divergence.

For this purpose we used the divergence derived directly from the CST. To
characterize the supergranulation scale, similarly to \cite{HGSD08} and
\cite{DET04}, we selected the low spatial resolution mode of the CST,
with a scale cut-off of 14 arcsec (around 10~Mm).  \cite{MTRR07} showed
that a scale cut-off of $10.2$~Mm is well adapted to the typical size
of supergranules.

The divergence patterns were tracked rigidly at the Carrington rate
of 13.2~\degr/day.  Then, the divergence fields were remapped onto
a latitude--longitude coordinate system using the Postel projection with a
pixel size of 0.21\degr, which is equivalent to a grid size of 2.55~Mm. The
time series spans seven days with a field of view centred on the equator
with the limits in latitude $(-28 \degr,+32 \degr)$ and in longitude $(290
\degr,330 \degr)$ in Carrington coordinates. The tracked region remains
visible on the disc and is located within about 30\degr of the limb.

\section{Properties of supergranules}
\subsection{Evolution of the divergence field}

We used the supergranular outflow core defined by \cite{DET04}. This is the
group of pixels of positive divergence where the local curvature of the function 
is negative, a segmentation criterion originally proposed by 
Strous \cite[see][]{RRRD07}.

In Figure~\ref{original1} each positive divergence is easily identifiable
in the segmented map. The segmentation processing was applied to all
the 336 divergence maps of our seven day sequence. Figure~\ref{segtime} shows
an example of the temporal evolution of the divergences and their segmented
counterparts during five~hours. In the selected field, three sunspots
are present as a divergence centre, like supergranules. However their contribution 
is negligible compared with the total number of detected divergences.
Figure~\ref{original2} displays a time--latitude slice of the divergence
(also from the segmented map) of the data cube, which exhibits the
temporal evolution of supergranule cores. The lengths of the persistent
flows are an indication of their lifetimes.

The proper determination of the lifetime of each cell requires 
cross-identifying of the supergranular centres between the consecutive
frames of the 3 D $(x,y,t)$ datacube.  For that purpose we used the time
labelling described \cite{RLRBM03}. This method was previously developed to track 
the evolution of solar granules, but has been fully adapted to  tracking 
supergranular cores.

In the seven day-long datacube with a time sampling of 30~min we identified
and tracked 4759 supergranular centres, which allowed us to follow
their properties during their evolution. This number is close to the
4977 supergranular lifetime histories followed by  \cite{HGSD08} over
a spatial window three times larger and a temporal window of five days which yielded
a 3D datacube 2.15 times larger. The similarity of these
numbers arises because our HMI/SDO data set has a higher
resolution than the MDI/SOHO  ~data~ used~ by  \cite{HGSD08}. We thus detect twice
as much supergranules which compensates for the smaller field of view
needed to use a seven-day series.

\begin{figure}
\includegraphics[width=9cm]{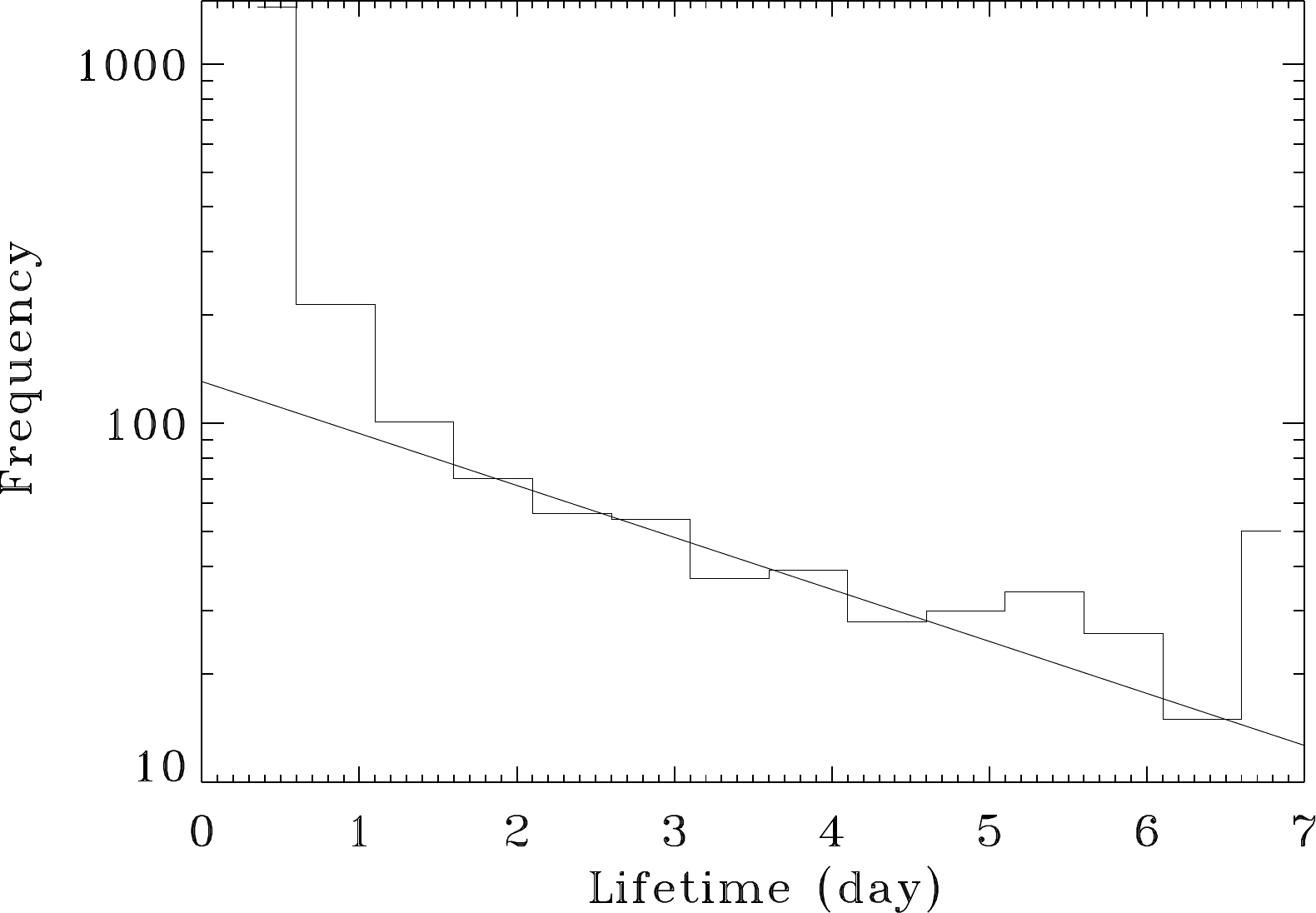}
\caption[]{Distribution of supergranular lifetimes for 3000 supergranules identified in
the temporal sequence, with a bin of 12~h. The solid line represents an exponential
fit to the feature-tracking data.}
\label{lifetime}
\end{figure}

The distribution of derived supergranular lifetimes is shown in
Figure~\ref{lifetime}.  For a temporal bin of 12 hours it may be
fitted with an exponential distribution in the form

\begin{equation}
N(t)=N(0)\exp[-t/\tau],
\end{equation}
where $\tau$ is the mean lifetime. The best-fit exponential function
(straight line in Figure~\ref{lifetime}) gives $\tau=1.5$ days, in
agreement with \cite{HGSD08}~ for a temporal bin of 12~h.

\subsection{Properties of supergranular cells}

\begin{figure*}
\sidecaption
\resizebox{4.0cm}{!}{\includegraphics{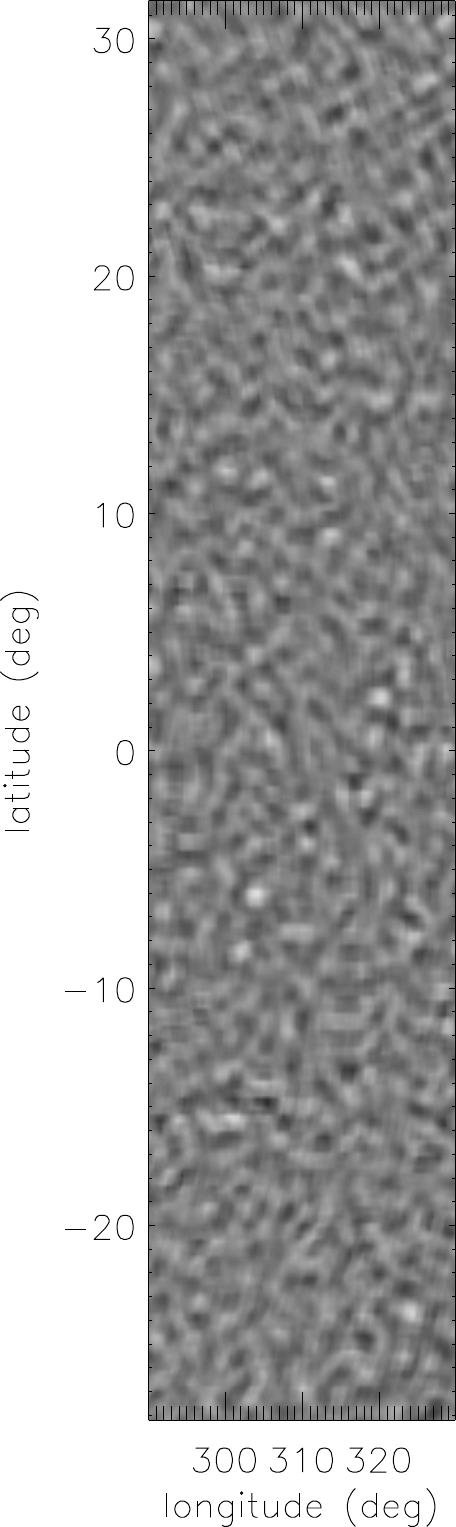}}
\resizebox{4.0cm}{!}{\includegraphics{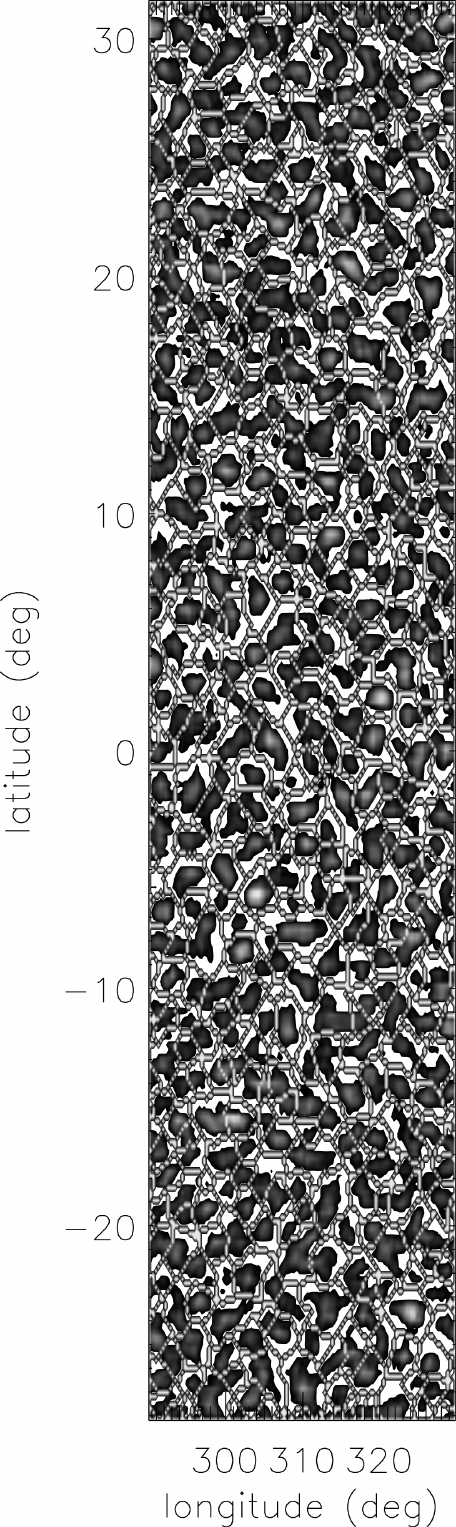}}
\resizebox{4.0cm}{!}{\includegraphics{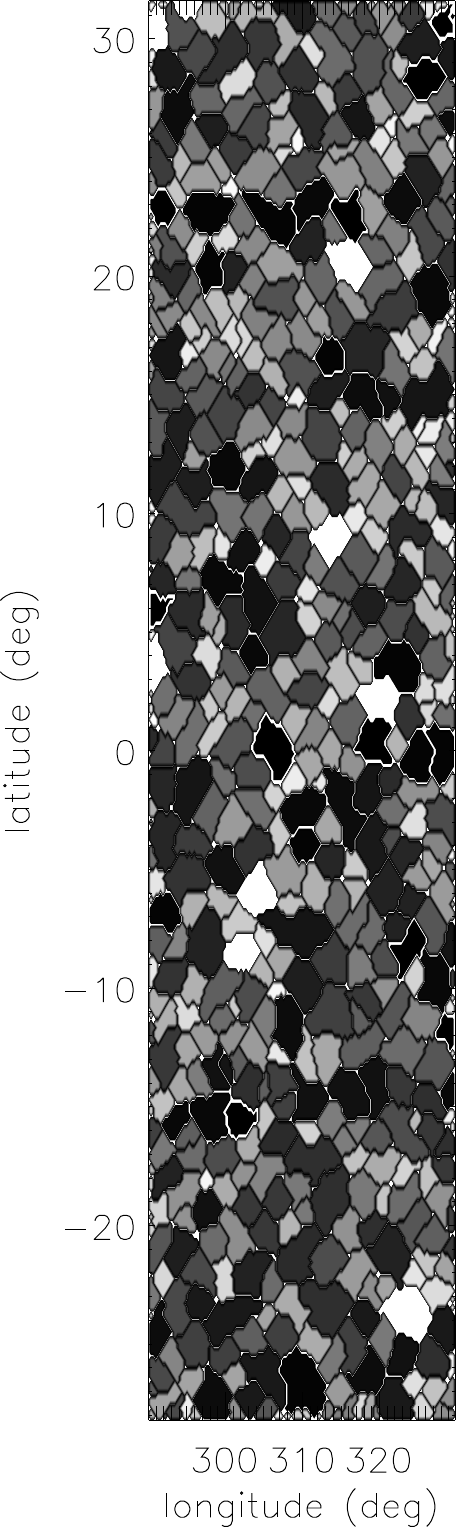}}
\caption[]{Supergranule outflow cores (divergences left) and the
associated watershed basin cells (white middle), and cell area
representation of the supergranules (right).}
\label{cells}
\end{figure*}

A more detailed investigation of the supergranulation is required
to properly identify each cell in the divergence map, following our
definition of one supergranule as a compact region of positive divergence
surrounded by lanes of the negative divergence. We detected the cells 
with a watershed basin segmentation applied to the divergence
maps, similarly to \cite{HGSD08}.  The divergence is treated as
a topographic map, with the divergence magnitude corresponding to the
height of various catchment basins (that are divergence cells). An
example of such a detection is displayed in Figure~\ref{cells}.  In that
plot we can clearly identify the cores of the cells (positive divergences) 
and the associated cells.

\begin{figure}
\resizebox{7.0cm}{!}{\includegraphics{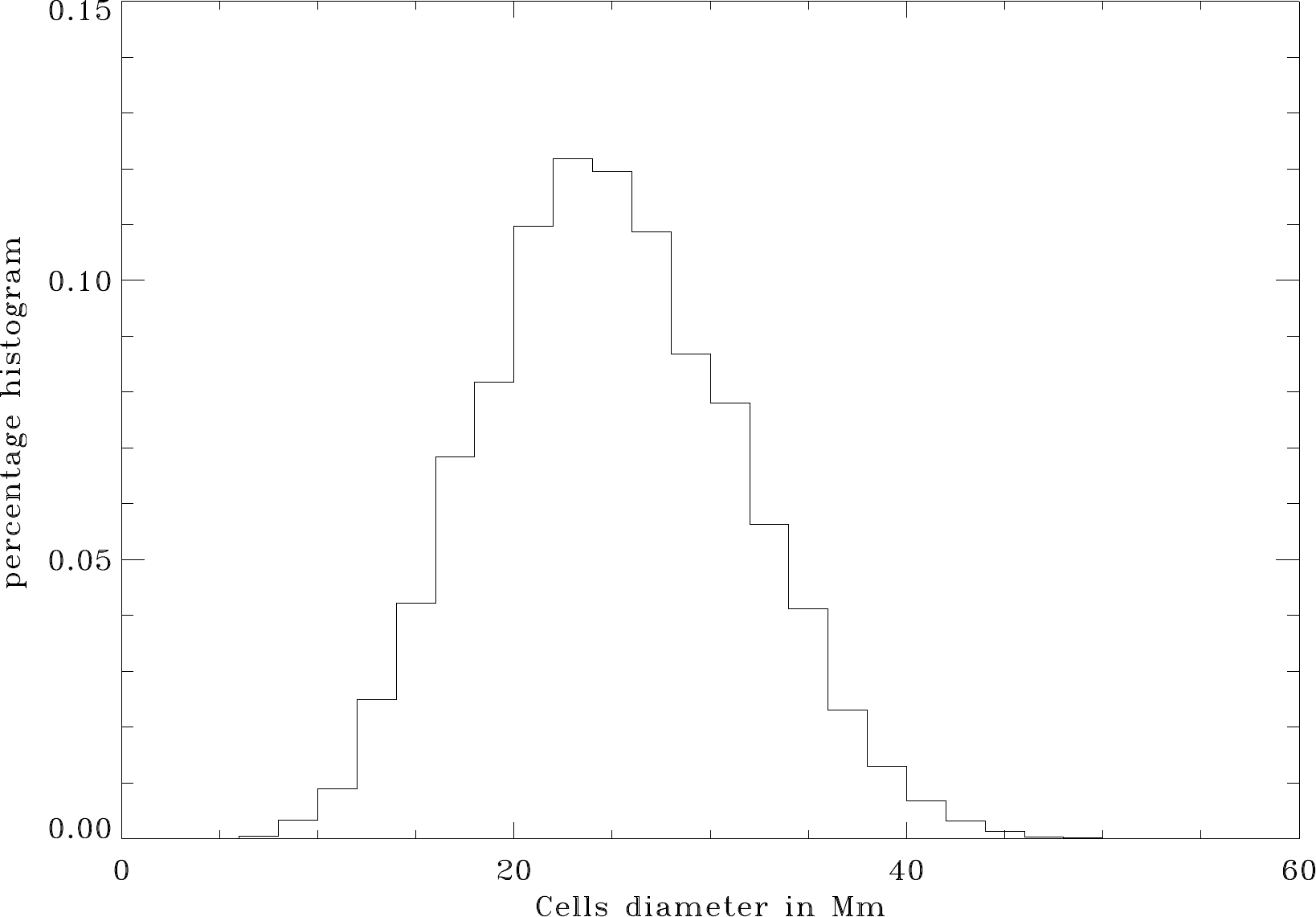}}
\caption[]{Number of supergranules versus their equivalent diameter.}
\label{histocells}
\end{figure}

The segmented cells allowed us to state additional statistical
properties. Figure~\ref{histocells} shows a histogram of a cell diameter that is 
peaked around 25~Mm  (${\rm diameter}=2\sqrt({\rm area}/\pi)$). Our resulting 
mean cell diameter is slightly smaller than found by \cite{HGSD08} -- 27.1~Mm -- and
\cite{DBDK04} -- 33~Mm --, because we detected smaller size
divergence centres by our processing. By combining the high spatial resolution of 
HMI/SDO data with the velocity computed by CST we were able to take into account 
all sizes of the divergence centres in the field of view by using watershed 
basin segmentation.

This result clearly demonstrates that the definition of a \emph{typical}
supergranule (size, diameter, lifetime, etc.) strongly depends on
the data set and the applied method. Therefore, it is very difficult
to properly describe \emph{a typical supergranular cell} when the method
does not take into account some of the representatives (typically
those ~with~ small~ sizes~ or ~weak ~divergence~ signals). To obtain proper 
physical characteristics of supergranulation the set of representatives must 
be complete. We~ are confident that~ this~ has~ been~ achieved in our analysis.

\begin{figure}
\includegraphics[width=9cm]{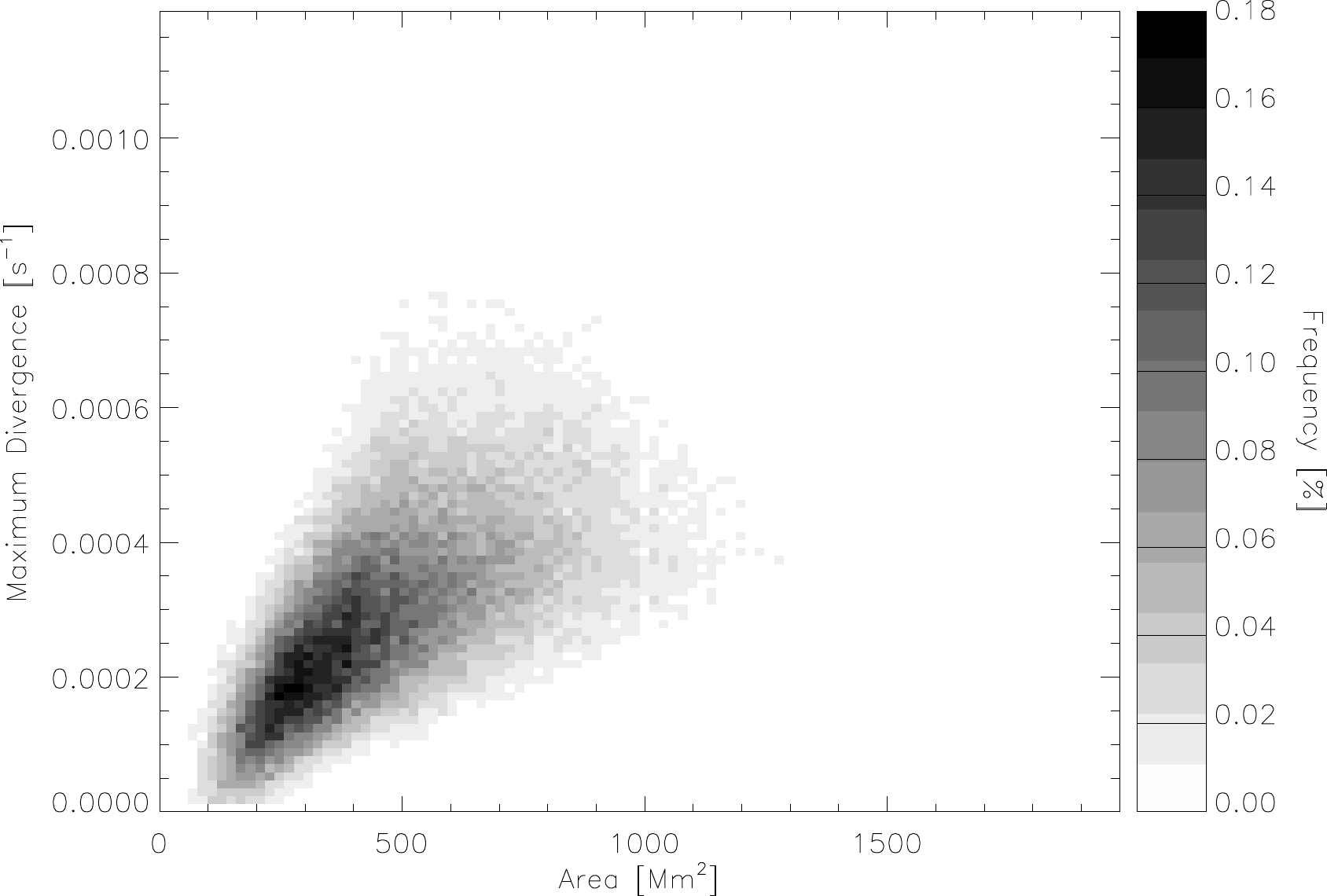}
\caption[]{Two-dimensional histogram of the maximum divergence versus
supergranular cell area.}
\label{aireflux}
\end{figure}

The relation between the area and maximum divergence reported
previously by \cite{HGSD08} and \cite{MTRR07} is confirmed by our
analysis. Figure~\ref{aireflux} shows this relation with an extension
toward low values. A linear relation appears to extend over the full
range of values with a very large scatter toward larger cells.

\begin{figure}
\includegraphics[width=9cm]{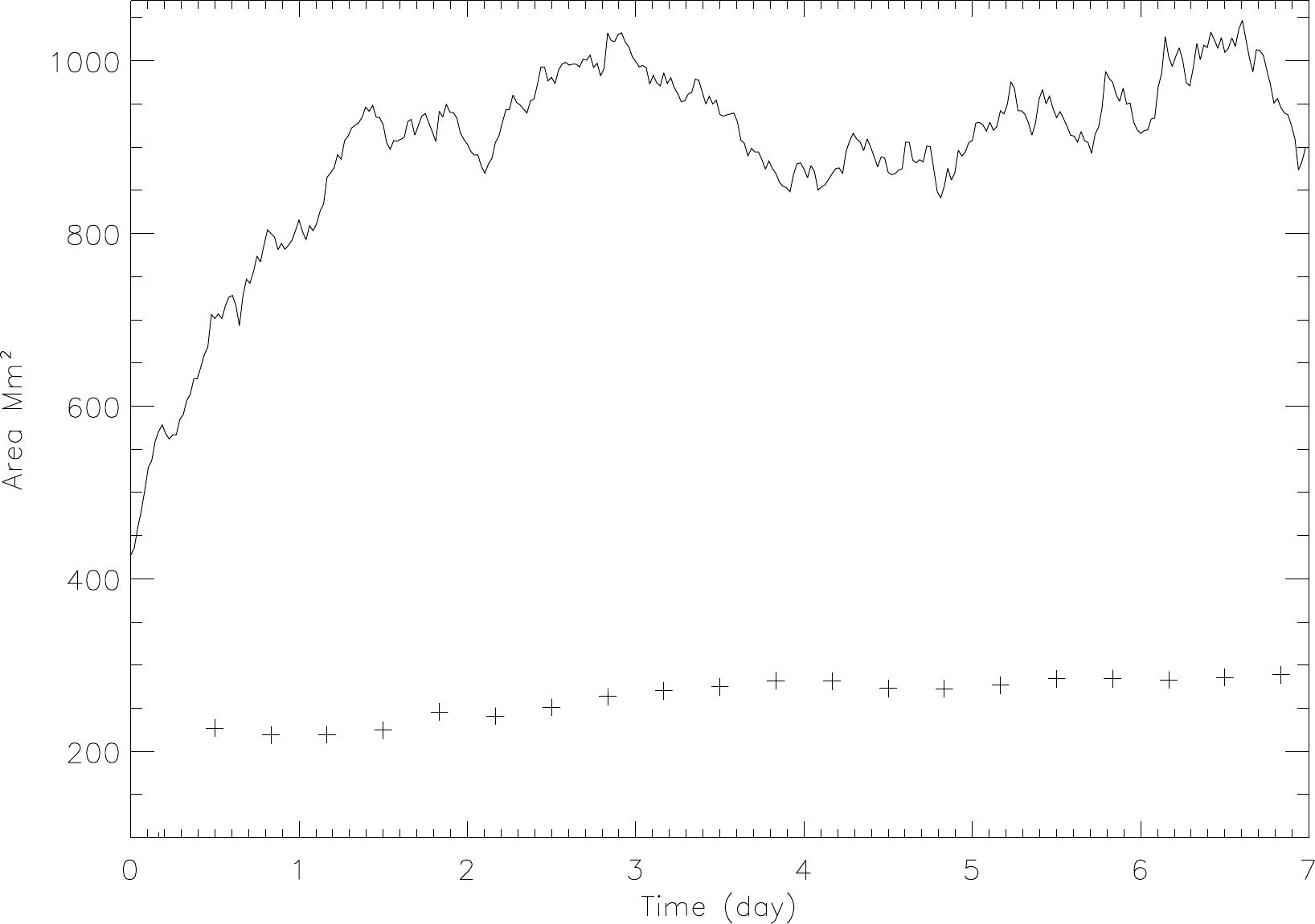}
\caption[]{Lifetime dependence of the area of the supergranule outflow core (cross) and 
lifetime dependence of the area of the 37 longest-living supergranules (solid line).}
\label{area_time2}
\end{figure}

\begin{figure}
\includegraphics[width=9cm]{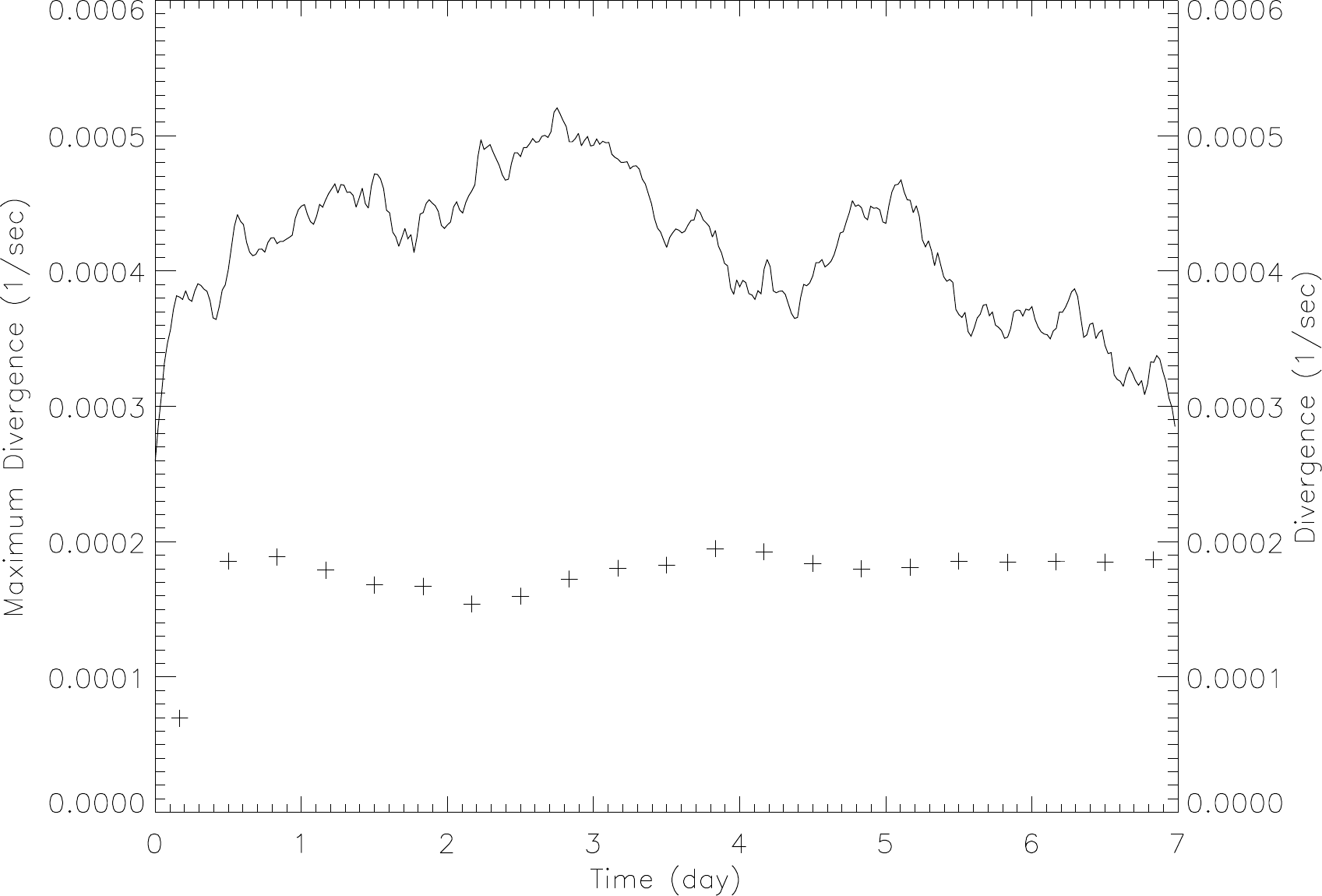}
\caption[]{Lifetime dependence of the maximun divergence of the supergranule outflow core (cross and 
Y coordinate axis left) and the temporal mean divergence of the 37 longest-living supergranules 
(solid line and Y coordinate axis right).}
\label{div_time2}
\end{figure}

The cross plots in Fig ~\ref{area_time2} and ~\ref{div_time2},
similar to Fig.~12 in \cite{HGSD08}, show the outflow core areas and
the maximum divergences averaged over each lifetime history and, additionally,
averaged over each bin of the histograms. We confirm the trend for larger
supergranules to live longer. However, we do not observe that supergranules
with larger divergences have longer lives -- the divergences appears to be 
quite constant regardless of the lifetime of the supergranule.

The solid lines in Fig~\ref{area_time2} and ~\ref{div_time2}
show the mean temporal evolution of the area and the divergence of the 37
longest-living supergranules. For the long-lived supergranules, we observe
an increasing area during the first 1.5~days with an area more or less
constant thereafter. The duration of such supergranules does
not allow us to see the final phase of their evolution. The divergence
intensity evolution shows an increasing phase during the first two or three
days followed by a slow decrease toward the end of our sequence.

\subsection{Advection of supergranules}
\begin{figure}
\resizebox{9.0cm}{!}{\includegraphics{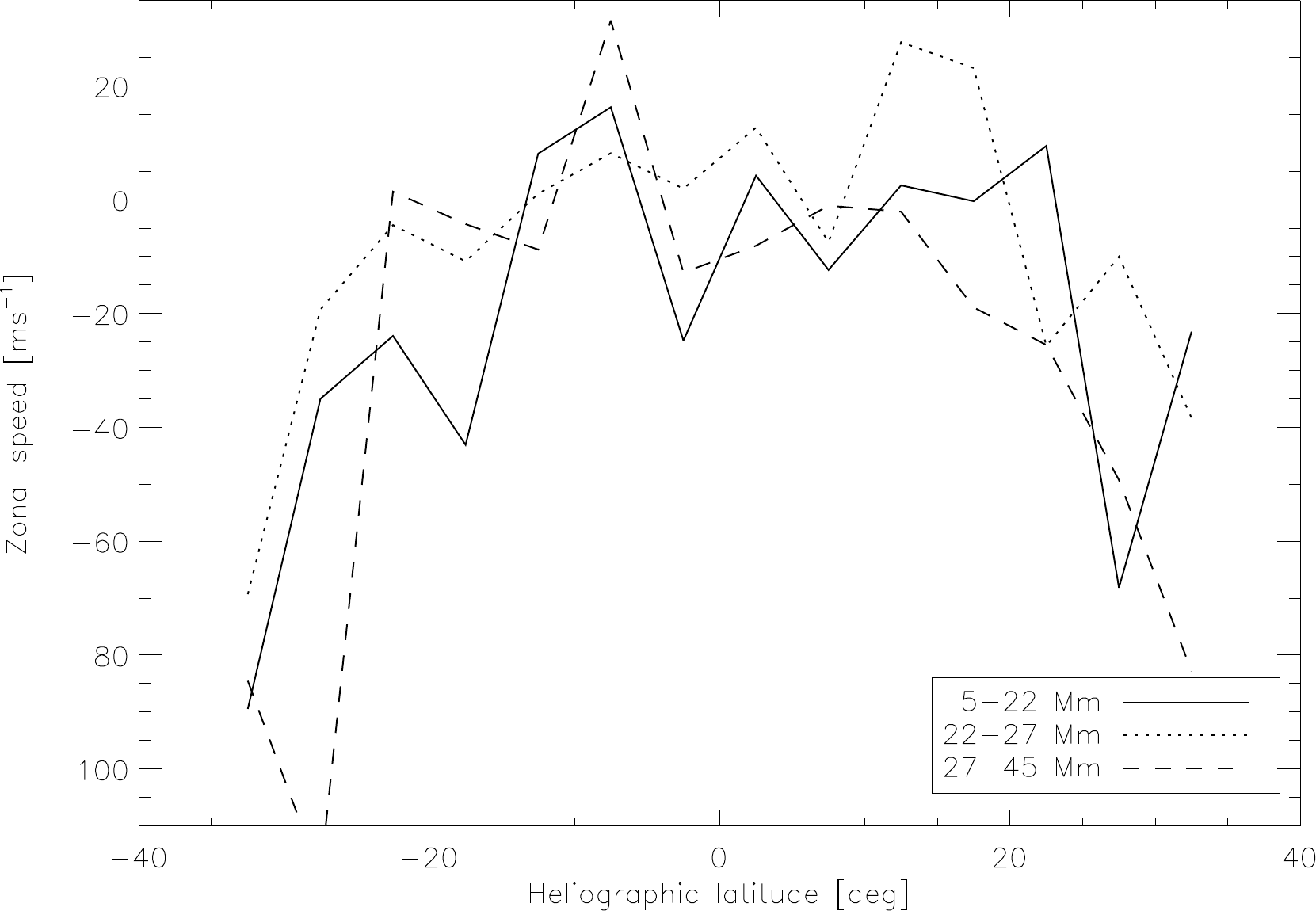}}\\
\resizebox{9.0cm}{!}{\includegraphics{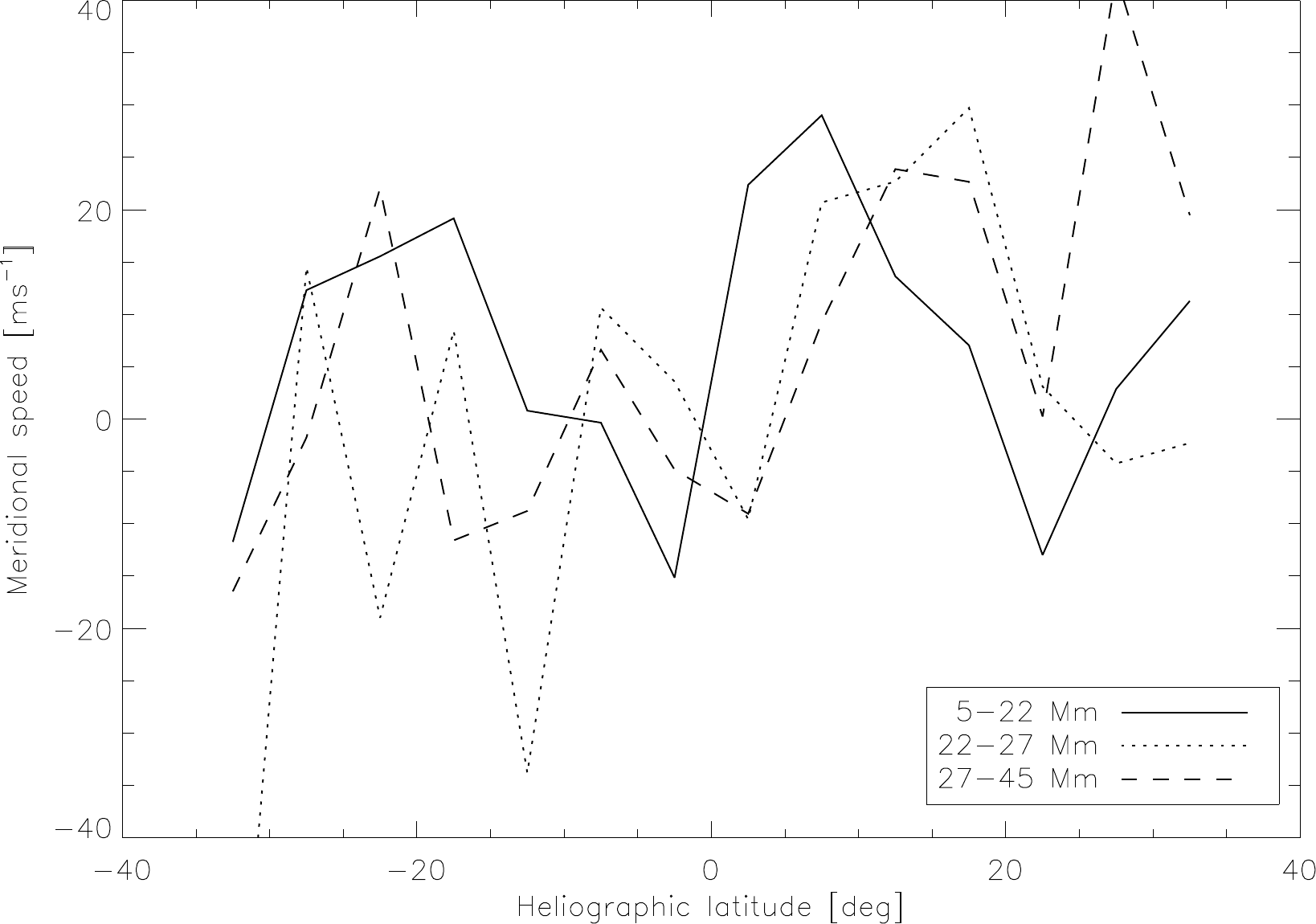}}
\caption[]{Zonal velocity component measured by tracking the supergranules cells (top). The differential 
rotation is clealy identified. The meridional component of the supergranule motion (bottom) shows 
a poleward motion, which does not depend on the size of supergranules.}
\label{velocity}
\end{figure}

It was also suggested by other authors
\citep[e.g.][]{SU90,SKS06,SKSB08,SKSKD09,Hat12} that supergranules
may be treated as objects that are carried by an underlying large-scale
velocity field, and hence may used as tracers for studies of such a velocity
field. Organised motions could serve as evidence for example of giant cells
\citep{Hat13}.  So far, most of these studies (if not all of them) were
driven by some versions of correlation techniques, which did not follow
individual supergranules, but rather \emph{a pattern} of supergranulation.
An additional filtering was used \citep[e.g. by][]{Hat12}
to select \emph{scales} that are to be tracked and hence discuss the
possible flow dependence on the size of these cells. One should proceed
with caution: without properly identifying the supergranular cells,
the scales selected by the filtering do not have to really isolate
only the supergranules with the appropriate size. Interpreting 
of such measurements may be rather complicated. \cite{Hat12} assumed
that the vertical extent of supergranular cells correspond to their
horizontal diameter and hence used the tracking of the supergranular
pattern with various scales to investigate the weak meridional
flow in depth. They concluded that the large supergranular cells (more
precisely a pattern with large scales) already ``feel'' the return
meridional flow some 50~Mm below the surface and hence move towards the 
equator. The hypothesis of the shallow meridional flow is supported by
a multicellular model of the meridional flow, which came recently from
local helioseismology \citep{Zhao13}.

In contrast to the traditional methodology, we may follow the individual
supergranular cells in time, and thus directly study the possible
dependence of the motions of supergranular cells on their sizes. A
drawback of this method is the significantly higher level of random noise,
which may be only decreased by using more supergranular histories and
averaging them, thus losing temporal resolution.

To collect a large statistical sample, the datacube covering seven days was split
into four chunks, each 42 hours long (84 frames). The splitting allowed
us to use larger fields of view undisturbed by edge effects than
if the datacube were used uninterrupted. Given the characteristic lifetime
of 1.5~days we assume that we have~ captured the evolution of most
of the supergranules in these chunks.  We did not consider supergranules
whose lifetime started or ended in the boundary frames; the history
of such supergranules would not have been complete.

Altogether we identified 14321 complete histories of individual
supergranular cells.  For each of these cell, its motion averaged over
its lifetime was computed by means of the difference of positions of
the gravity centre in the first frame, where the cell was detected,
and the frame, where the cell was detected the last time. Because of the
very discrete coverage of the field of view, we averaged the motions of
supergranules over the bands of 5\degr in latitude.

The ensemble of supergranule results was split into three bins according
to their effective size. The boundary values were chosen so that the
bins were more or less equally populated. In this case, there were 3005
supergranules in the bin of 5--22 Mm, 5744 supergranules in the bin of
22--27 Mm and 5556 cells in the bin of 27--45 Mm.

Figure~\ref{velocity} shows the resulting velocity components additionally 
averaged in longitude. The noise level of each plotted point is
estimated to be 10 m/s. Since the reference frame for tracking the
frames in the datacube is the Carrington rotation rate, the differential
rotation is clearly visible in the zonal component (upper panel). In
the meridional component (lower panel) the trend of weak single-cell
poleward meridional flow is visible.

Our results contradict those derived by \cite{Hat12}. He reported (1)
the decreasing amplitude of the poleward meridional flow with the size
(more precisely wavelength) of supergranules, the flow eventually turned 
equatorward for supergranules with wavelengths longer than 50~Mm, and (2)
the increasing amplitude of the zonal flow component with increasing size
of supergranules. Our analysis shows that both zonal and meridional flows
are independentof the supergranules size. 

\cite{Hat12} claimed that the onset of the meridional flow reversal occurs 
for supergranule wavelengths longer than 50~Mm. Interestingly, in our sample 
of more than 14\,000 supergranules with complete histories, there is only one 
supergranule longer than 50~Mm. A similar histogram by \cite{HGSD08}
points to a similar conclusion. The question therefore is what structures
are represented by Fourier wavelengths of 50~Mm: obviously,
it cannot be individual supergranules. 

A possible improvement of this approach would be to use more supergranular
cells as tracers, which would reduce the noise level and also allow for
a better resolution in both latitude and size of supergranules.

\subsection{Effect of the magnetic field}

\begin{figure}
\includegraphics[width=9cm]{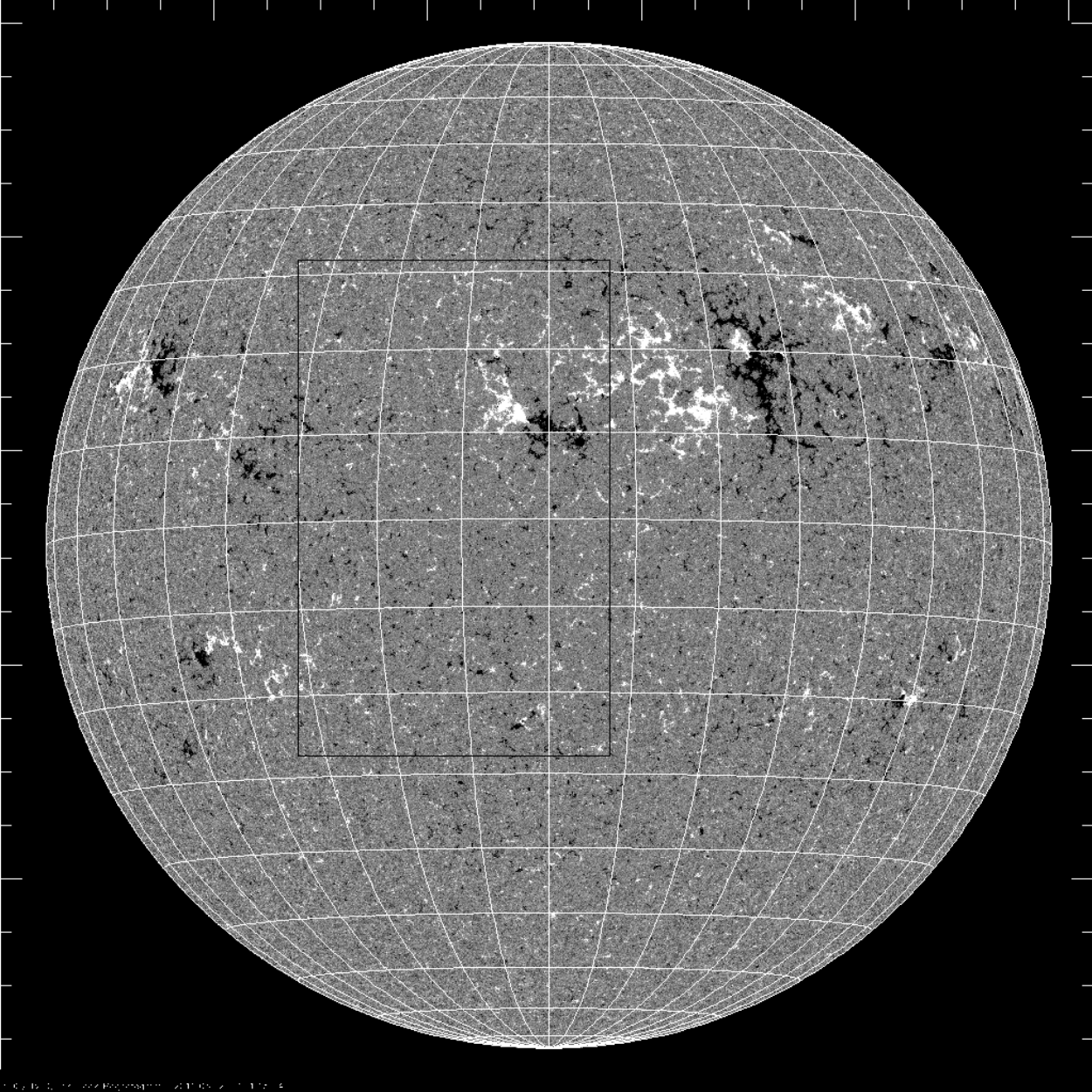}
\caption[]{Map of the longitudinal magnetic field on 12 May 2011 showing the latitude of the
concentrated magnetic activity.}
\label{Magnetic}
\end{figure}

\begin{figure}
\resizebox{4.0cm}{!}{\includegraphics{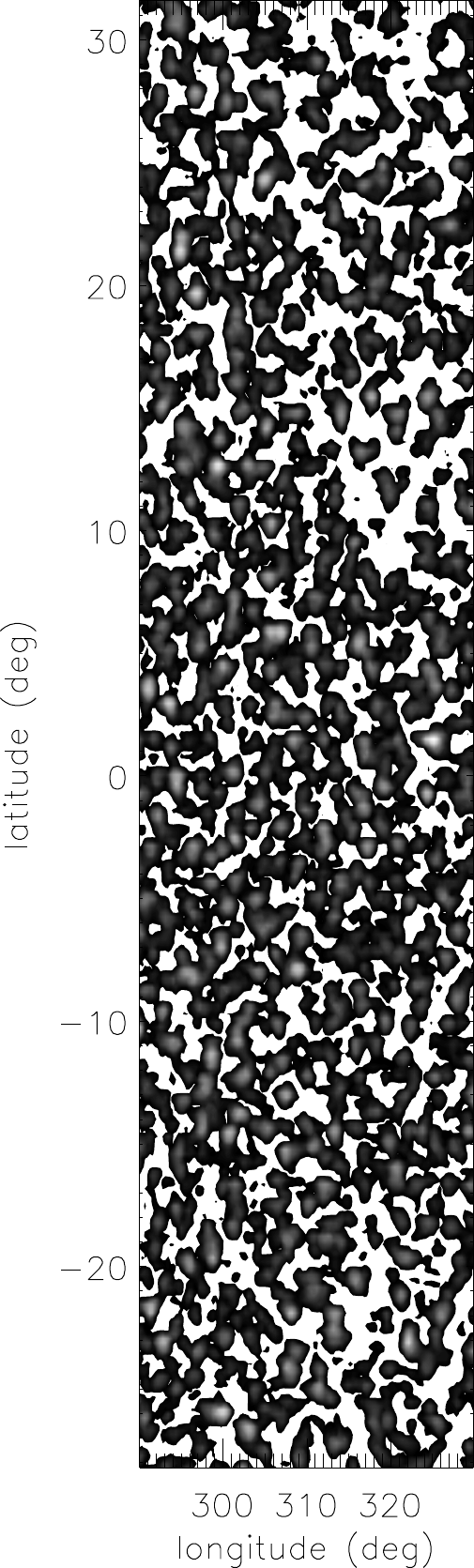}}
\resizebox{4.0cm}{!}{\includegraphics{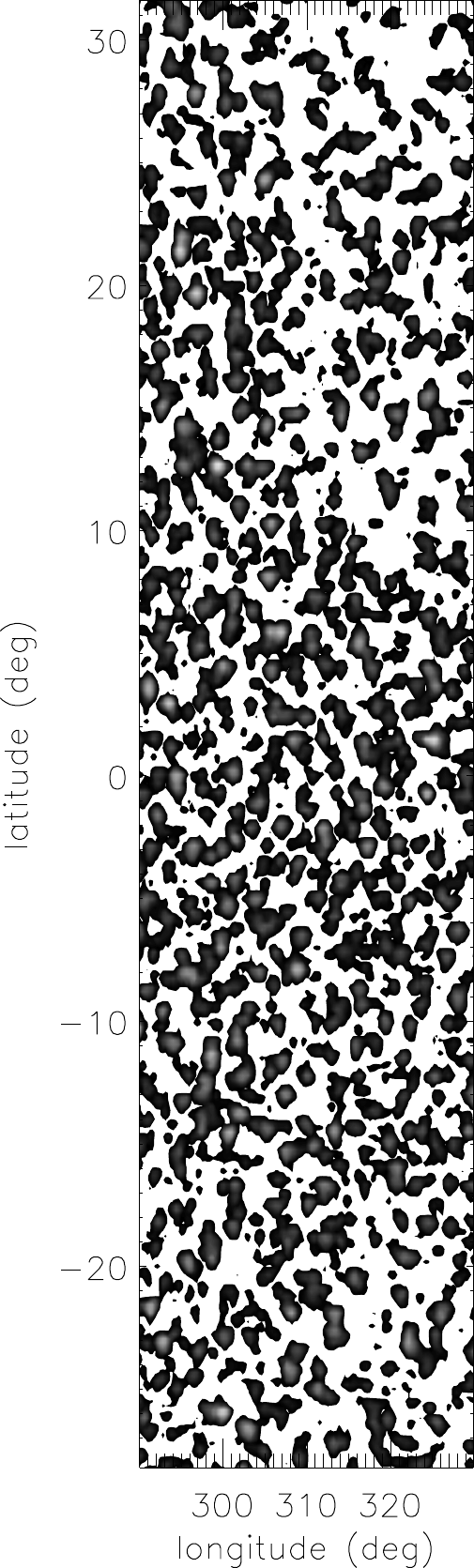}}
\caption[]{Divergence (larger than 0.0002876~s$^{-1}$ in each frame) integrated over 7 days of observations (left).
Identical plot with values higher than 0.000342~s$^{-1}$ over the 7 days (right). 
We note in the right upper corner a lower density of the stronger divergence due to the magnetic activity
at that location during the 7 days of the sequence. We also note the same density below a latitude
$-20$\degr in the bottom right corner, but the magnetic field was only active for 2 days.}
\label{ampli}
\end{figure}

\begin{figure}
\includegraphics[width=9cm]{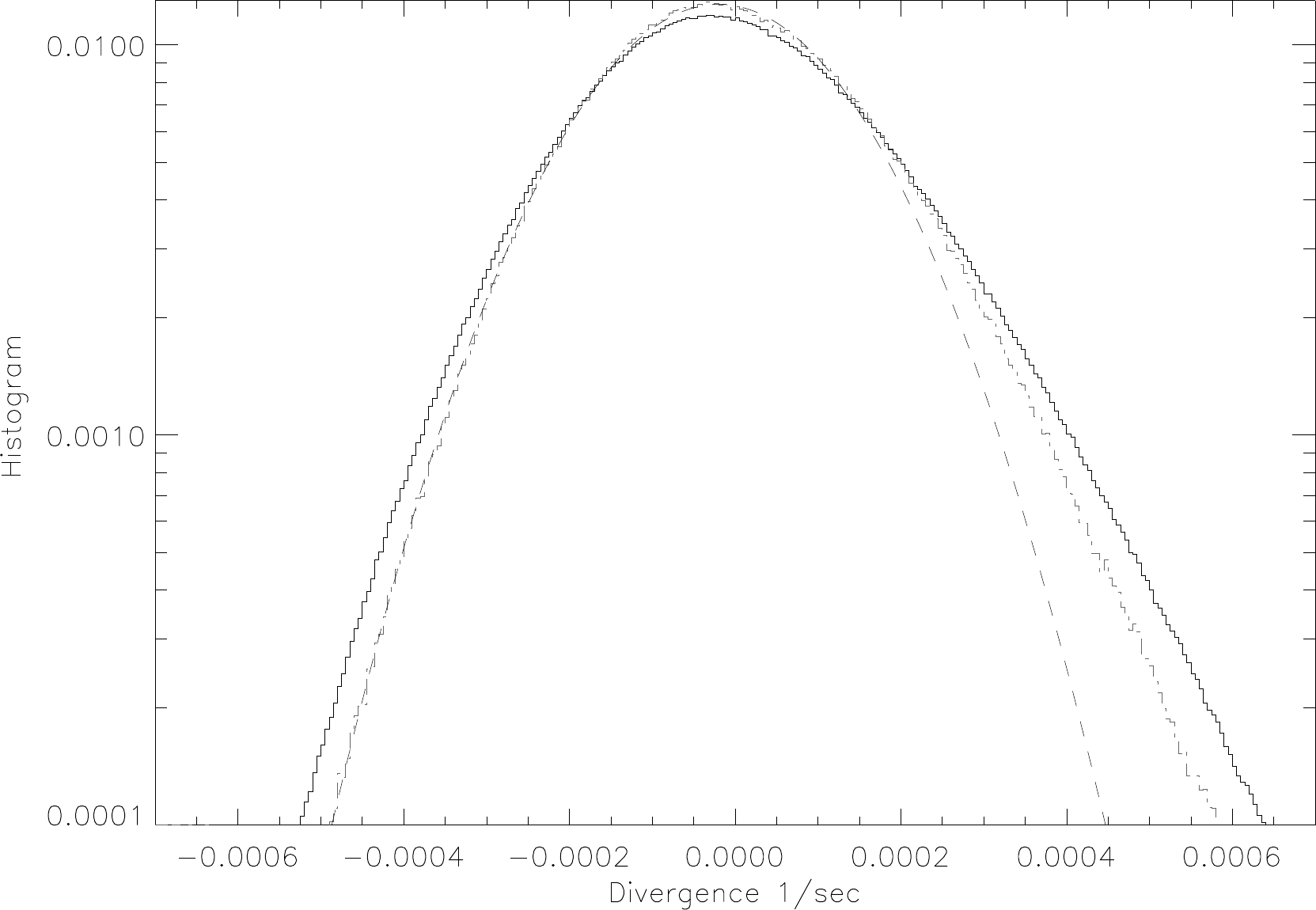}
\caption[]{Comparison of the divergence distribution in the entire field of view (solid line) and
in the upper right corner (dashed dotted), where a strong magnetic field is present. A Gaussian distribution 
with the same standard deviation is overplotted for comparison (dashed line). }
\label{compar}
\end{figure}

It is well established today that magnetic fields~ affect
the dynamic of the solar convection. Figure~\ref{Magnetic} shows the
magnetic context in field studied (dark rectangle) on May 12, 2011.
The magnetic field is located in the top right part of our field of view. 
To test the possible influence of that magnetic activity on the
amplitude of the divergence, we took thresholded divergences integrated
over time defined as

\begin{equation}
\langle D(x,y)\rangle=\int\limits_{\rm day 0}^{\rm day 7}{\rm d}t\, d(x,y,t),
\end{equation}
where $d(x,y,t)=\bnabla_{\rm h}\cdot\bvec{v}(x,y,t)$ for ($\bnabla_{\rm
h}\cdot\bvec{v})(x,y,t) > d_{\rm thresh}$ and $d(x,y,t)=0$ otherwise.

Figure~\ref{ampli} shows $\langle D(x,y)\rangle$ with a threshold $d_{\rm
thresh}=0.0002876$~s$^{-1}$ and an identical plot with a threshold $d_{\rm
thresh}=0.000342$~s$^{-1}$, corresponding to 2.5 and 3.0
$\sigma$ of the divergence fluctuations. Note that yhe lower density
of the stronger divergence is  centered on the upper right corner because of the
magnetic activity at that location. This is confirmed by the histogram in
Figure~\ref{compar}, where the divergence distribution in the magnetic
region is more concentrated at small amplitudes than is the full field-of-view 
divergence distribution. These distibutions
clearly show a larger positive wing than that~ for~ the ~respective~
Gaussian ~ distribution~ with ~the~ same ~standard-deviation. We note
also below the latitude $-20\degr$ in the bottom right corner the same
smaller density, but the magnetic field was only active 2 days during
our time sequence.

\subsection{Relation of the magnetic network with divergence amplitude}

\begin{figure}
\includegraphics[width=9cm]{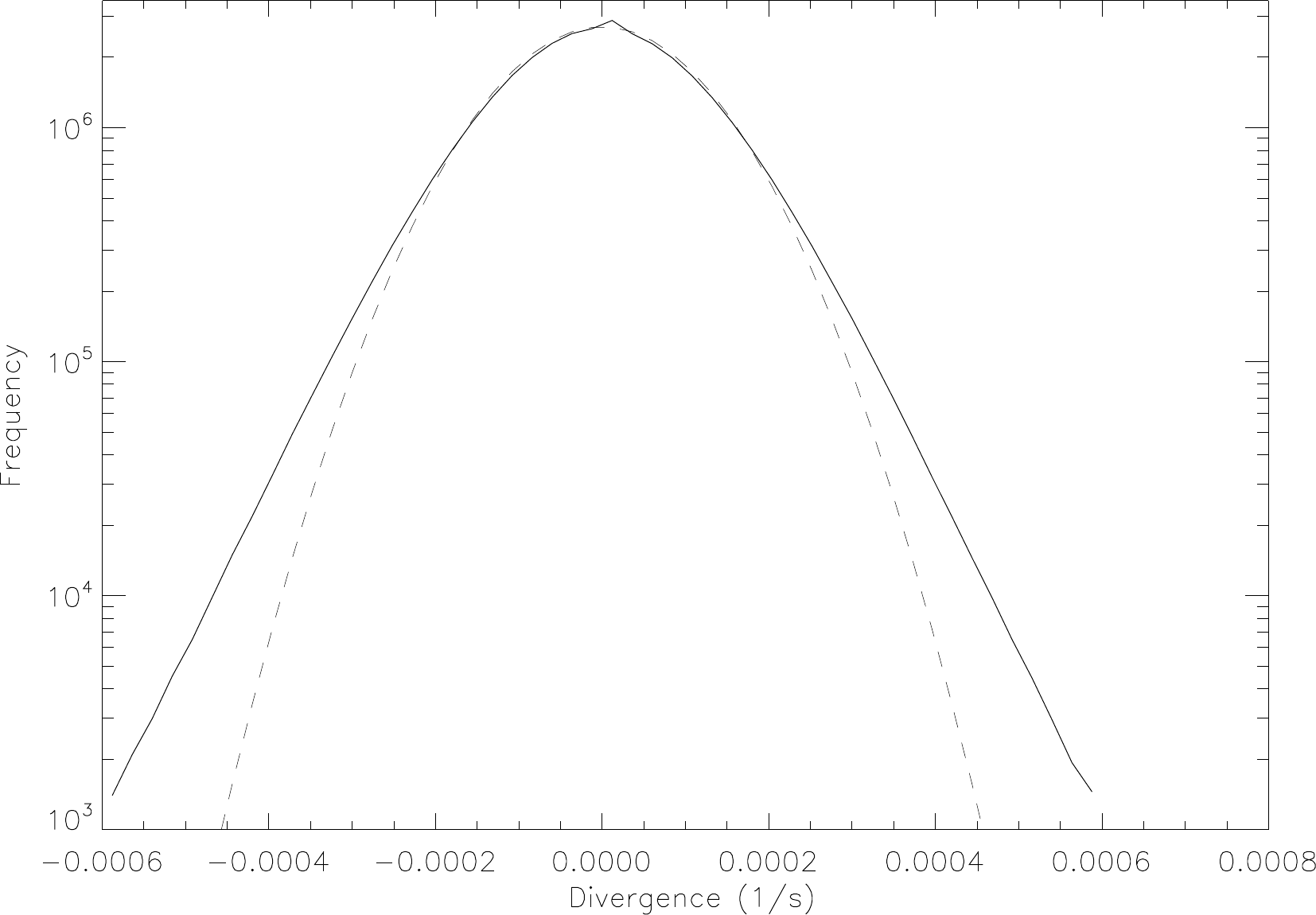}
\caption[]{Histogram of the divergences during the first 48h of the time sequence. 
A Gaussian distribution with the same standard deviation is overplotted for comparison (dashed line).}
\label{histodiv}
\end{figure}

The magnetic network is known to be located at the edge
of the supergranules. We studied the contribution of the large
amplitude divergences (supergranules) to building the network. First,
we used the hypothesis that the motions of supergranules accross the solar
surface are purely horizontal. From the velocities
measured by CST $v_x$ and $v_y$, we computed the longitudinal $v_\phi$
and latitudinal $v_\theta$ components. These velocities were projected
with a grid with a regular angle spacing of 0.2047 degrees in both $\phi$
and $\theta$. From the projected $v_\phi$ and $v_\theta$ we computed the divergence
field. Figure~\ref{histodiv} displays the divergence. This distibution shows wings 
larger than  that~ for~the~ corresponding~ Gaussian~, which~ reflects~ 
the~ intermittency ~of~the~ supergranulation~ \citep{RMRRBP08}. 

\section{Conclusions}

By following the proper motions of the solar granules, we can
revisit the dynamics of the solar surface at high spatial and temporal
resolutions from hours to months and years with the SDO data. Applying
the CST to HMI/SDO white-light images gives direct access to
the solar plasma motion on the solar surface. The long continuous seven 
day series offers the opportunity of studying the
properties of supergranulation and its evolution.  The CST code and the
multi-resolution technique directly produce the divergence field,which minimizes
the error propagation during the scaling. The spatial window of
10.2~Mm is well adapted to the typical size of supergranules; thus we can use 
the divergence patterns as proxies of supergranules. We confirm most of the 
results described by \cite{HGSD08}, except for the divergence increase relative to 
the increasing lifetime. This is probably due to the different ways of calibrating  
the divergence amplitude.

We found the mean diameter of supergranular cells to be about 25~Mm,
which is smaller than found previously.  The reason might be a better 
detection of the small-scale divergences. We also observe a smaller 
amplitude of the divergence in the region of strong magnetic
activity, which probably has some influence on the diffusion of the
magnetic elements in this region.

Applying the tracking on the postel projected divergence highlights the 
solar differential rotation and a poleward circulation trend
of the meridian flow regardless of the size of supergranules.
This result does not confirm the different behaviour of the largest cells
found by \cite{Hat12}, where the poleward meridional flow disappears.
This trend must be confirmed with a longer sequence (one month at least)
to reduce the noise, but this suggests that such a data analysis can be used
to study the meridional flow with a smaller temporal window than for other methods. 
Improvements are expected with a longer data-set analysis
on the meridional flow and its evolution during the solar cycle.

\begin{acknowledgements}
We thank the HMI team members for their hard work. We thank the
German Data Center for SDO for providing SDO/HMI data.
This work was granted access to the HPC resources of CALMIP under 
the allocation 2011-[P1115]. M.~\v{S} is supported by the Czech Science 
Foundation (grant P209/12/P568). Tato pr\'ace vznikla s podporou na dlouhodob\'y 
koncep\v{c}n\'\i{} rozvoj v\'yzkumn\'e organizace RVO:67985815 
a v\'yzkumn\'eho z\'am\v{e}ru MSM0021620860. We would like to give special 
thanks to the anonymous referee for his or her helpful comments and recommendations.

\end{acknowledgements}

\end{document}